\documentstyle[fises,multicol,epsfig]{fisestex}

\begin{document}

\title{
Perturbative continued-fraction method for weakly interacting Brownian
spins and dipoles
}
\author{
J. L. Garc\'{\i}a-Palacios
}
\address{
Dept. de Teor\'{\i}a y Simulaci\'on de Sistemas Complejos,\cite{chaingang} \\
Instituto de Ciencia de Materiales de Arag\'on,
C.S.I.C. -- Universidad de Zaragoza, E-50009 Zaragoza, Spain
}

\maketitle

\begin{multicols}{2}
\narrowtext

The continued-fraction (CF) method was developed systematically by
Risken and co-workers to solve problems of arbitrary fluctuations in
nonlinear systems\cite{risken}.
However, this efficient technique is limited to problems with a few
variables,\cite{hantalbor90} which in practice means systems of
noninteracting entities (particles, spins, etc.)
In this communication, we illustrate how to extend the CF method
to weakly {\em coupled} systems with the problem of relaxation in
classical spins.

The Fokker--Planck equation governing the $T\neq0$ dynamics of a
classical spin $\vec{s}$ is given by\cite{bro63}
\[
2\tau_{{\rm D}}
\frac{\partial W}{\partial t}
=
-\frac{\partial{}}{\partial\vec{s}}
\cdot
\left\{
\frac{1}{\lambda}
\vec{s}\wedge\vec{B}
-
\vec{s}
\wedge
\left[
\vec{s}
\wedge
\left(
\vec{B}
-
\frac{\partial {}}{\partial\vec{s}}
\right)
\right]
\right\}
W
\;,
\]
where $\lambda$ is the dissipation parameter, $\vec{B}=-(\partial{\cal
H}/\partial\vec{s})$ the effective field, and $\tau_{{\rm
D}}\propto T^{-1}$ the free diffusion time.
If we write this equation as $\partial W/\partial t={\cal L}_{\rm FP}
W$, the dynamical equation for any average $\left\langle
f\right\rangle$ is given in terms of the adjoint Fokker--Planck
operator by ${\rm d}\left\langle f\right\rangle/{\rm d} t=\langle{\cal
L}_{\rm FP}^{\dagger}f\rangle$.

However, the underlying Landau--Lifshitz damping term
$-\lambda\vec{s}\wedge\big(\vec{s}\wedge\vec{B}\big)$ is nonlinear
in $\vec{s}$ and couples the equations for the moments.
To handle the infinite systems of coupled equations it is convenient
to introduce the spherical harmonics $X_{l}^{m} = e^{{\rm i} m\varphi}
P_{l}^{m}(s_{z})$, and take advantage of their recurrence relations.
Writing ${\cal L}_{\rm FP}^{\dagger}$ in terms of the rotation
operator $\vec{{\cal J}}= -{\rm
i}\,\vec{s}\wedge(\partial{}/\partial\vec{s})$
\begin{equation}
\label{Ladjoint:J}
{\cal L}_{\rm FP}^{\dagger}
=
-\case{{\rm i}}{\lambda}
\vec{B}
\cdot
\big[
\vec{{\cal J}}
+
\lambda
\big(
\vec{s}\wedge\vec{{\cal J}}
\big)
\big]
-
\vec{{\cal J}}^{2}
\;,
\end{equation}
we get the equations for the $X_{l}^{m}$ in the form of a recurrence
relation that can be solved by CF methods.

When the spins {\em interact}, the structure of ${\cal L}_{\rm
FP}^{\dagger}$ is the same with the field created by the other spins
$\vec{b}_{\rm d}$ included in $\vec{B}$.
This allows to separate the free evolution and interaction
terms, ${\cal L}_{\rm FP}^{\dagger} = {\cal L}_{{\rm FP},0}^{\dagger} +
\vec{b}_{\rm d} \cdot \vec{{\cal J}}_{\rm LL} $, where $\vec{{\cal
J}}_{\rm LL}$ is the operator multiplying $\vec{B}$ in Eq.\
(\ref{Ladjoint:J}).
For weak coupling, we can expand the averages in the coupling
parameter $\left\langle f\right\rangle=\left\langle
f\right\rangle_{0}+\left\langle f\right\rangle_{1}+\cdots$, getting
a hierarchy of equations
\begin{eqnarray*}
{\rm d}\left\langle f\right\rangle_{0}/{\rm d} t
&=&
\big\langle {\cal L}_{{\rm FP},0}^{\dagger}\, f\big\rangle_{0}
\\
{\rm d}\left\langle f\right\rangle_{1}/{\rm d} t
&=&
\big\langle {\cal L}_{{\rm FP},0}^{\dagger}\, f\big\rangle_{1}
+
\big\langle \vec{b}_{\rm d}\cdot\vec{{\cal J}}_{\rm LL}\, f\big\rangle_{0}
\\
{\rm d}\left\langle f\right\rangle_{2}/{\rm d} t
&=&
\big\langle {\cal L}_{{\rm FP},0}^{\dagger}\, f\big\rangle_{2}
+
\big\langle \vec{b}_{\rm d}\cdot\vec{{\cal J}}_{\rm LL}\, f\big\rangle_{1}
\\[-1.5ex]
&\vdots&
\end{eqnarray*}
To illustrate our procedure, let us denote by $X_{\ell}^{m}$ and
$Y_{p}^{q}$ the spherical harmonics of two of the spins, and consider
the equation for $X_{\ell}^{m}$ (i.e., $f=X_{\ell}^{m}$ above).
Since the field created by the second spin depends on $Y_{p}^{q}$, one
needs $\left\langle X_{\ell}^{m}\,Y_{p}^{q}\right\rangle_{i}$,
$i=0,1$, which could be obtained writing the above equations also for
$f=X_{\ell}^{m}\,Y_{p}^{q}$.
This leads to another complication, since $\vec{b}_{\rm
d}\cdot\vec{{\cal J}}_{\rm LL}\, f$ then involves products of three
spherical harmonics.
However, $Y_{1}^{\mu}Y_{p}^{q}$ can be expanded as a linear
combination of the $Y_{p}^{q}$'s, restoring a structure solvable by CF
methods.

\begin{figure}
\begin{center}
\epsfig{file=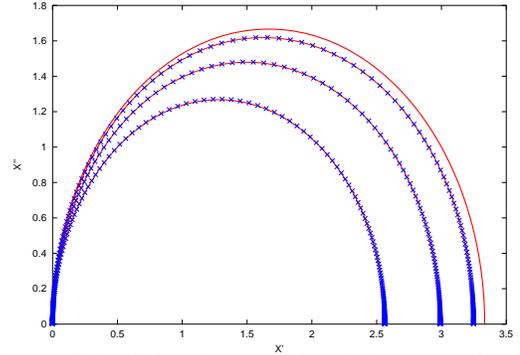, width=2.7in}
\caption[]{Cole--Cole plots of the dielectric polarisability
(imaginary vs. real part) for various couplings $\varrho$. Lines:
Zwanzig formula; crosses: perturbative CF results.}
\label{fig1}
\end{center}
\end{figure}

These manipulations result in a complicated algorithm, so it is
convenient to check it out against some solvable case.
Fortunately, for freely rotating spins the problem reduces to that of
electric dipoles, where the perturbative calculation can be done
analytically, yielding\cite{zwa63}
\[
\frac{\alpha(\omega)}{\alpha_{0}(\omega)}
=
1
\!-\!
\frac{\alpha^{2}\varrho^{2}}{1\!+\!{\rm i}\omega\tau_{{\rm D}}}
\left\{
{\cal R}
\left[
1
\!+\!
\frac{{\rm i}\omega\tau_{{\rm D}}}{2(4\!+\!{\rm i}\omega\tau_{{\rm D}})}
\right]
{}-
\frac{{\cal S}}{1\!+\!{\rm i}\omega\tau_{{\rm D}}}
\right\}
\]
where $\alpha_{0}(\omega)$ and $\alpha$ ($\propto T^{-1}$) are the
dynamic and static polarisabilities of noninteracting dipoles,
$\varrho$ is the dipole density, and ${\cal R}$ and ${\cal S}$ are
certain ``lattice sums'' of the interaction tensor.
Figure \ref{fig1} shows that our perturbative CF approach gives
accurately the analytical results.
Therefore, this approach extends the applicability of the efficient CF
techniques to (weakly) interacting systems, and allows one to handle
problems involving {\em driven} systems, evolving in {\em nonlinear
potentials}, {\em coupled}, and subjected to {\em dissipation} and
{\em fluctuations}

The author acknowledges financial support from DGES (PB98-1592) and F. Falo for useful discussions.

\end{multicols}

\end{document}